\documentclass[twocolumn,english,aps,prl,groupedaddress,superscriptaddress]{revtex4}

\usepackage{graphicx,epsfig,units}
\usepackage{amsmath,amsfonts,mathrsfs,amsbsy,bm,babel}
\usepackage{dcolumn}
\usepackage{bm}
\usepackage[utf8x]{inputenc}
\usepackage{rotating}
\usepackage{array}
\newcolumntype{L}[1]{>{\raggedright\let\newline\\\arraybackslash\hspace{0pt}}m{#1}}
\newcolumntype{C}[1]{>{\centering\let\newline\\\arraybackslash\hspace{0pt}}m{#1}}
\newcolumntype{R}[1]{>{\raggedleft\let\newline\\\arraybackslash\hspace{0pt}}m{#1}}
\begin{document}
\title{Structure and magnetic properties of {\it Ln}MnSbO ({\it Ln} = La and Ce) and CeMnAsO}
\author{Qiang Zhang}\email{qzhangemail@gmail.com}
\affiliation{Ames Laboratory, Ames, IA, 50011, USA}
\affiliation{Department of Physics and Astronomy, Iowa State University, Ames, IA, 50011, USA}
\author {C. M. N. Kumar}\affiliation{J\"{u}lich Centre for Neutron Science (JCNS), Forschungszentrum J\"{u}lich GmbH, Outstation at SNS, Oak Ridge National Laboratory, Oak Ridge, Tennessee 37831, United States}
 \affiliation{Chemical and Engineering Materials Division, Oak Ridge National Laboratory, Oak Ridge, Tennessee 37831, United States}
\author{Wei Tian}
\affiliation{Quantum Condensed Matter Division, Oak Ridge National Laboratory,Oak Ridge,Tennessee 37831, USA}
\author{Kevin W. Dennis}
\affiliation{Ames Laboratory, Ames, IA, 50011, USA}
\author{Alan I. Goldman}
\affiliation{Ames Laboratory, Ames, IA, 50011, USA}
\affiliation{Department of Physics and Astronomy, Iowa State University, Ames, IA, 50011, USA}
\author{David Vaknin}\email{vaknin@ameslab.gov}
\affiliation{Ames Laboratory, Ames, IA, 50011, USA}
\affiliation{Department of Physics and Astronomy, Iowa State University, Ames, IA, 50011, USA}

\date{\today}

\begin{abstract}
Neutron powder diffraction (NPD) study of \textit{Ln}MnSbO (\textit{Ln }$=$ La or Ce) reveals
differences between the magnetic ground state of the two compounds due to
the strong Ce-Mn coupling compared to La-Mn. The two compounds adopt the
\textit{P4/nmm} space group down to 2 K and whereas magnetization measurements do not
show obvious anomaly at high temperatures, NPD reveals a C-type
antiferromagnetic (AFM) order below $T_{\mathrm{N}} = 255 $ K for LaMnSbO
and 240 K for CeMnSbO. While the magnetic structure of LaMnSbO is preserved
to base temperature, a sharp transition at $T_{\mathrm{SR}}  = 4.5 $K is observed in
CeMnSbO due to a spin-reorientation (SR) transition of the
Mn$^{\mathrm{2+}}$ magnetic moments from pointing along the $c$-axis to the \textit{ab}-plane. The
SR transition in CeMnSbO is accompanied by a simultaneous long-range AFM
ordering of the Ce moments which indicates that the Mn SR transition is
driven by the Ce-Mn coupling. The ordered moments are found to be somewhat smaller
than those expected for Mn$^{\mathrm{2+}}$ ($S = 5/2$) in insulators, but
large enough to suggest that these compounds belong to the class of
local-moment antiferromagnets. The lower $T_{\mathrm{N\thinspace }}$ found in
these two compounds compared to the As-based counterparts ($T_{\mathrm{N}} = 317$ for LaMnAsO, $T_{\mathrm{N}} = 347$ K
for CeMnAsO) indicates that the Mn-$Pn$ ($Pn=$ As or Sb) hybridization that
mediates the superexchange Mn-$Pn$-Mn coupling is weaker for the Sb-based compounds.

\end{abstract}
 \maketitle
\section{INTRODUCTION}
Spin and orbital degrees of freedom of transition-metal ($T = $ Fe, Co, Mn, ...) pnictide-based compounds such as A$T_2$As$_2$ ($A=$ Ba, Sr, Ca, ...)
or $RT$AsO ($R =$ rare-earth elements)  settle into distinct ordered magnetic ground states that for some, especially the Fe based pnictides, slight doping or external pressure suppresses the static magnetic structure and induces superconductivity\cite{Johnston2010,Lumsden2010,Dai2015}. In all these systems, the transition metal atoms form a square lattice (or slightly distorted into a rectangular lattice) with a corrugated layer of nearest neighbor (NN) pnictides  ($Pn =$ P, As, Sb, Bi) that mediate extended super-exchange coupling among the transition metals.
The ground state for the parent ``1111" and ``122"  Fe-based pnictides is almost without exception a spin-stripe
antiferromagnetic (AFM) plane with varying inter-planar stacking that depends on the details of the elements separating the
FeAs planes\cite{Huang2008,delaCruz2008}. On the other hand, the Co based pnictides tend to form ferromagnetic (FM) planes, static
(i.e., CaCo$_{1.86}$As$_2$)\cite{Quirinale2013} or dynamic (i.e., SrCo$_2$As$_2$)\cite{Jayasekara2013}.
It is by now understood that both Fe- and Co-based ``122" family of pnictides compounds exhibit relatively strong AFM next-NN (NNN)
exchange coupling ($J_2$) that, due to its competition with the NN coupling ($J_1$; $J_1 \sim J_2 $), leads to the stripe-like
magnetic structure or to fluctuations with the same motif.  The relatively strong $J_1$ is intimately related to
strong hybridization of $p$- and $d$-orbitals of $Pn$ and $T$, respectively\cite{Ma2008,Haule2009,Lee2010,Yildirim2009}.
By contrast, Mn-based ``1111" pnictides $Ln$MnAsO ($Ln=$ La, Nd, Ce) tend to undergo a simple checkerboard (C-type) AFM
structure\cite{Marcinkova2010,Emery2011,Zhang2015} with much higher transition temperatures than their Fe-based counterparts indicating effectively stronger NN interaction i.e., $J_1 > 2J_2$. Neutron powder diffraction (NPD) study of the isostructural PrMnSbO has shown similar C-type Mn magnetic ordering  below $T_{\mathrm{N}}\approx 230$ K,
followed by a spin reorientation (SR) to the $ab$-plane\cite{Kimber2010}. In addition,  a tetragonal to orthorhombic transition at low temperatures was
found at $\sim35$ K presumably due to the Pr $4f-$electron degree of freedom. It is interesting to note that this variable magnetic behavior of
transition metal pnictides (i.e., $AT_2Pn_2$  or $RTPn$O) is strikingly different than that of other insulating  transition metal oxides. For instance, the Li orthophsphate family, Li$T$PO$_4$ ($T=$ Mn, Fe, Co, and Ni), all  exhibit the same AFM ground state that differs only in the direction
of the ordered moment, and the moment size to a good degree of accuracy obeys classical Hund's rules under crystal field effect as a local moment\cite{Santoro1966,Vaknin1999,Toft-Petersen2012}.  In fact, the average moment size for transition metal pnictides tends to be significantly smaller than the local moment in insulators suggesting a high degree of itineracy in the $d$ shell and an instability of the electronic structure.

Here, we report on the preparation, structure and the magnetic properties of LaMnSbO and CeMnSbO by employing neutron diffraction techniques and bulk magnetization measurements.
Room temperature structures of these two compounds have been reported \cite{Schellenberge2008}.
Although both systems are expected to exhibit similar properties in the MnSb triple-
plane, the presence of Ce or other magnetic rare-earth element has been shown to have an effect on the nuclear structure and the magnetic structure of the transition metal\cite{Kimber2010,Marcinkova2010,Zhang2015,Zhang2013}. These Sb-based pnictides are also
instructive in examining systematically the effect of the of As-Sb replacement in $Ln$Mn$Pn$O.

\section{EXPERIMENTAL DETAILS}
The synthesis of the polycrystalline LaMnSbO compound is similar to that reported by  Schellenberge et al.\cite{Schellenberge2008}. Two
starting materials MnO and LaSb were mixed thoroughly in stoichiometric proportions (LaSb was prepared firstly by reacting La and
Sb powders at 600 $^{\rm o}$C for 5 h and subsequently at 950 $^{\rm o}$C for 12 h under 1/3 atm. of argon).
The mixed powder was  sealed in a tantalum tube under argon at 1/3 atm. and sintered at 1120 $^{\rm o}$C for 24 h.  The CeMnSbO was prepared in one step using Ce, Sb, and MnO as starting materials by mixing  them thoroughly, sealing the mixture in a quartz tube under 1/3 atm of argon and sintering at 1120 $^{\rm o}$C for 24 h.  All mixing and thorough grinding of the starting materials were conducted under inert environment (i.e., glove box under argon) to minimize metals oxidation.
Magnetization measurements were carried out in a  superconducting quantum  interference device (Quantum Design MPMS-7S, SQUID) magnetometer.
Neutron powder diffraction (NPD) measurements on  $\approx 4$ g LaMnSbO and CeMnSbO samples
were conducted on the HB1A  triple-axis spectrometer with a fixed-incident-energy 14.6 meV (located at the High Flux Isotope Reactor, HFIR, at the Oak Ridge National Laboratory, USA). The measurements on HB1A were performed with an HOPG analyzer
to yield a lower background scattering (providing approximately 1 meV energy resolution). Two HOPG crystals were used to filter out the $\lambda/2$ component from the incident beam. The data between
$2 <T<300$ K of LnMnSbO (La and Ce) were collected using an  {\it orange} cryostat, whereas a high temperature furnace was used for the measurements for CeMnSbO between $300<T<670 $ K.

Neutron powder diffraction (NPD) measurements were also performed on the time-of-flight powder diffractometer, POWGEN, located at the Spallation Neutron Source at Oak Ridge National Laboratory. The data were collected with neutrons of central wavelengths 1.333~{\AA} and 3.731~{\AA}, covering the $d$-spacing range $0.4 - 6$  and $1.65-14$ {\AA}, respectively. About 2.5~g samples were loaded in a vanadium container of 8~mm diameter and measured in an orange cryostat in the temperature range of $2-300$ K. All  the neutron diffraction data were analyzed using Rietveld refinement program Fullprof suite\cite{Fullprof}.

\section{RESULTS AND DISCUSSION}
\begin{figure} \centering \includegraphics [width = 0.72\linewidth] {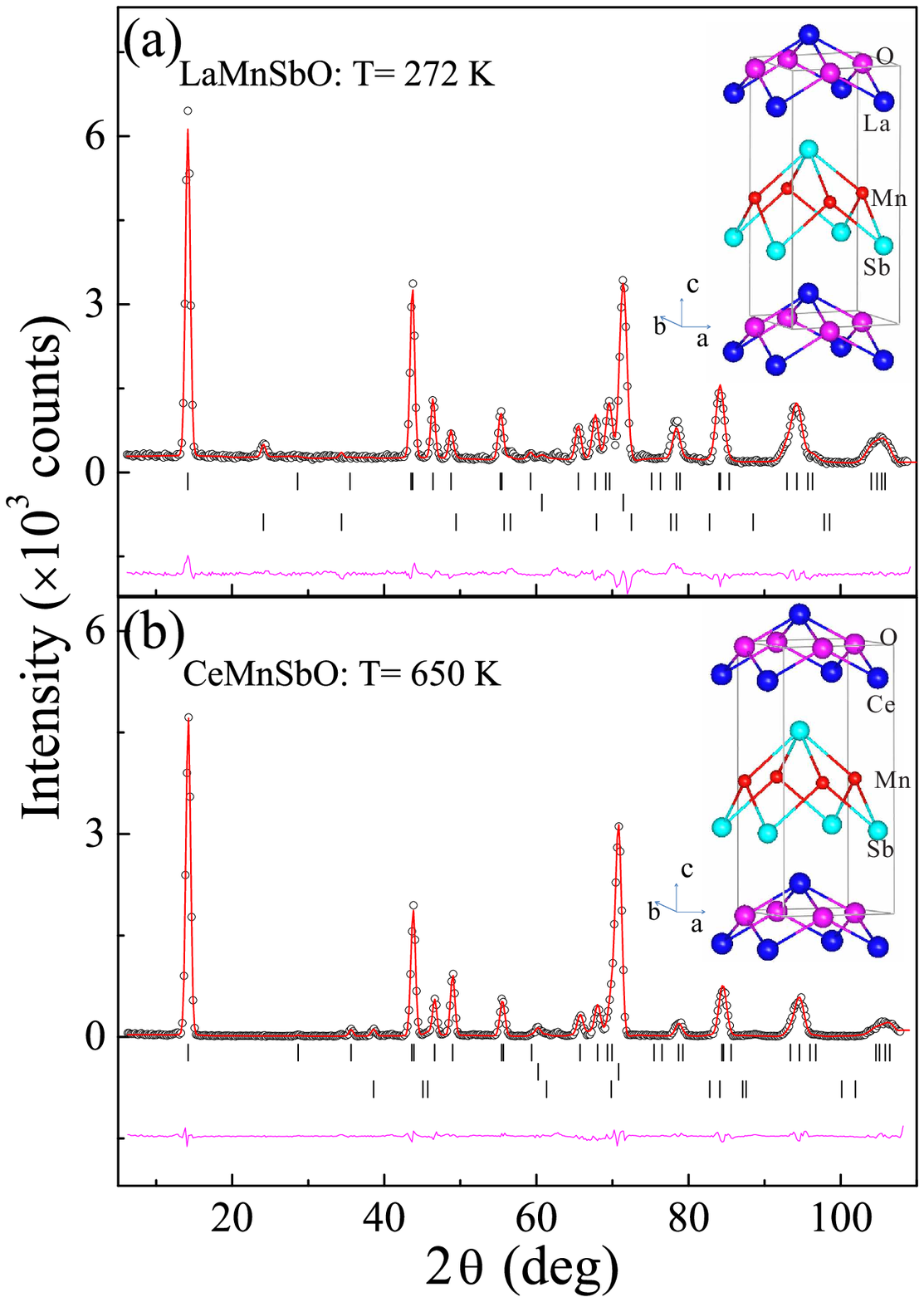}
\caption{(color online)  Neutron diffraction patterns  measured on HB1A
  (a) at 272 K for LaMnSbO and (b) at 650 K for CeMnSbO. The insets shows
  the corresponding graphic  representation of the crystal structure. The blue spheres in (a) and (b) indicate La and Ce atoms, respectively.The observed data
and the Rietveld fit are indicated by the open circles and
solid lines, respectively, with the difference curve at the bottom. The
vertical bars mark the positions of Bragg reflections for the expected "ZrCuSiAs''-type structure (top),
Al sample holder  and impurity phase of Sb in LaMnSbO and MnSb in CeMnSbO (bottom).}
\label{fig:Structure}
\end{figure}
\begin{figure} \centering \includegraphics [width = 1\linewidth] {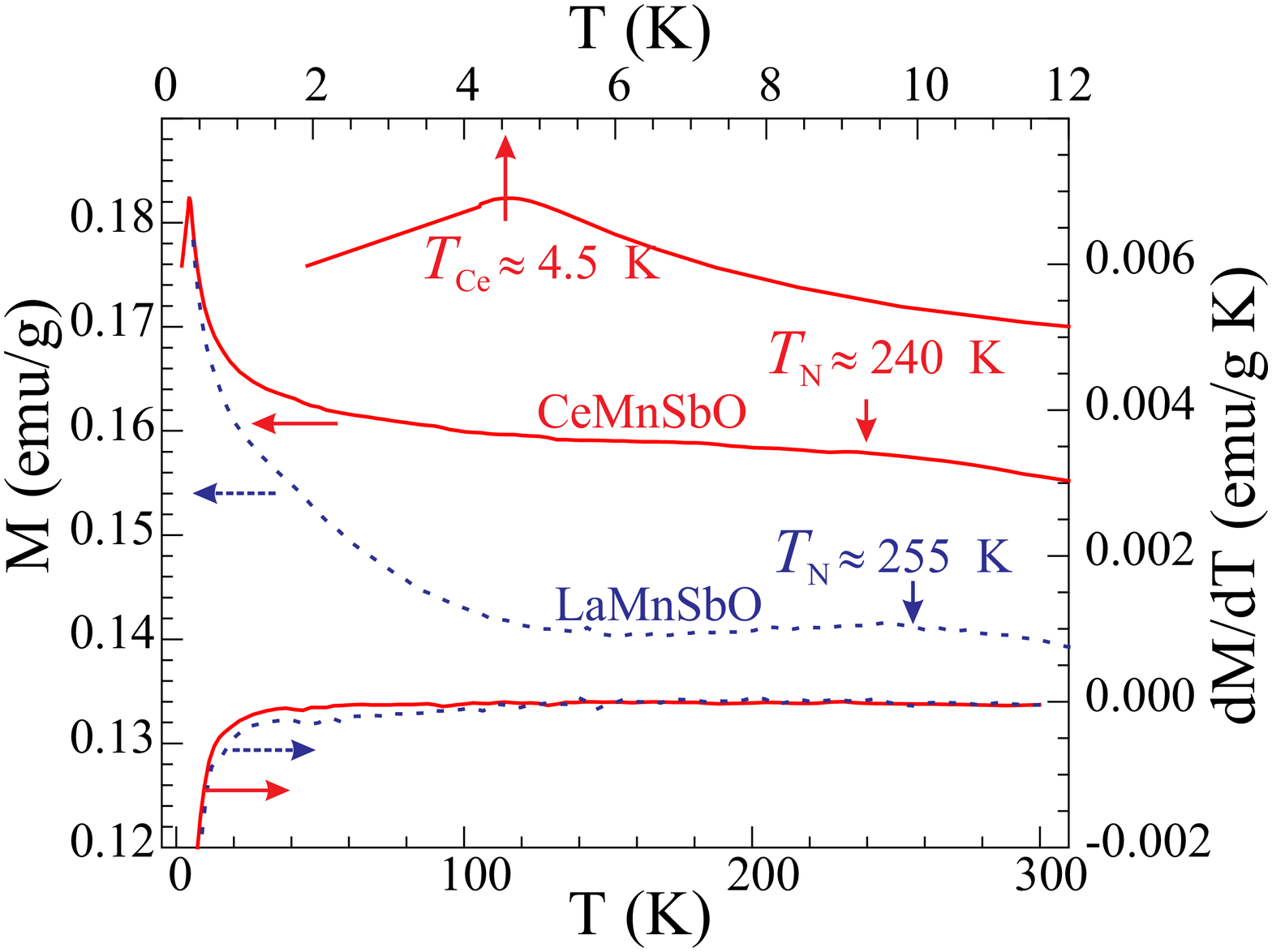}
\caption{(color online)   Temperature dependence of the
magnetization of LaMnSbO and CeMnSbO at applied magnetic of 0.01 T. The two lowest curves (dashed and solid lines) show the derivatives of the magnetization for both samples with respect to temperature.  Note that  both curves do not show anomalies at the AFM transition of Mn at around 250 K, however a sharp peak associated with the Mn spin-reorientation and Ce ordering is observed at $T_{\rm SR} = T_{\rm CE }\approx 4.5$ K.  }
\label{fig:Mag}
\end{figure}
Figure \ \ref{fig:Structure} shows neutron powder diffraction scans of LaMnSbO and of CeMnSbO  measured on HB1A  above their AFM transition temperatures.  The solid lines are best calculated intensities based on the ZrCuSiAs-type $P4/nmm$ space group consistent with previous powder x-ray diffraction reports\ \cite{Schellenberge2008} (note that the measurement and refinement include peaks from an Al sample can). The refinement of the CeMnSbO indicates that there is a small amount of MnSb present in the sample as an impurity phase.  Both LaMnSbO and CeMnSbO  maintain tetragonal $P4/nmm$ symmetry to the lowest temperature, namely 2 K as summarized in Table\ \ref{tab:Refined_Parameters}.
\begin{figure} \centering \includegraphics [width = 0.72\linewidth] {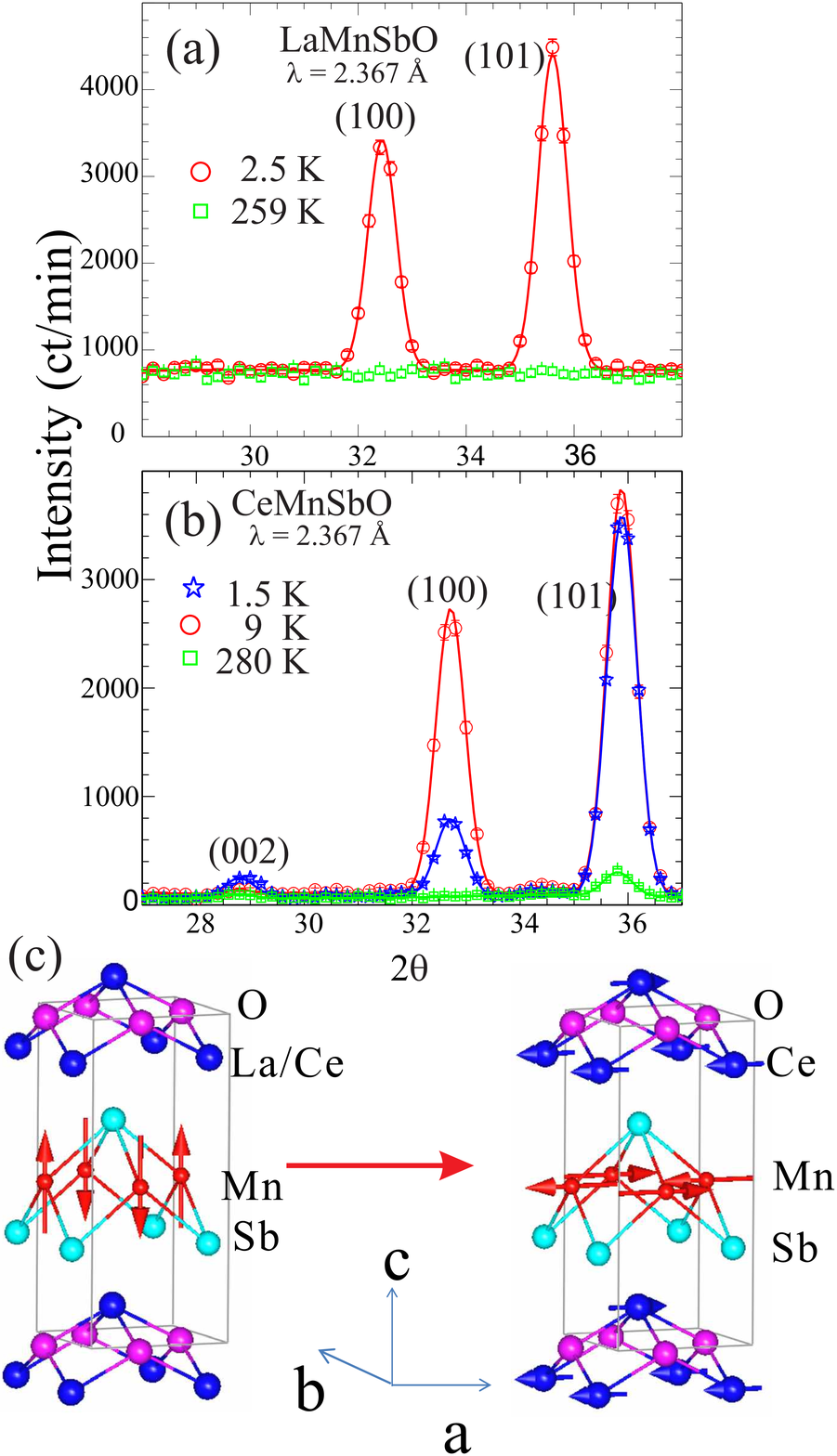}
\caption{(color online) (a) Low angle neutron diffraction scans showing the emergence of the (100) and the increase in intensity (101) reflections for (a) LaMnSbO and (b) CeMnSbO due to a simple Mn AFM ordering (checkerboard-like) as shown in the left part of (c). For CeMnSbO, as the temperature is lowered further (below $\approx 4.5$ K) the (100) Bragg reflection is significantly reduced in intensity and simultaneously the  (200) gains intensity, indicating a uniform spin-flop of the Mn moments from pointing along the $c$-axis to pointing in the $ab$-plane and the emergence of Ce ordering with a magnetic structure as depicted in the right side of panel (c). }
\label{fig:LowTpeaks}
\end{figure}
Three-dimensional (3D) graphic representation of the crystals, based on our Rietveld analysis, are included as insets in Fig.\ \ref{fig:Structure}.
Figure\ \ref{fig:Mag} shows the magnetization measurements  of LaMnSbO and CeMnSbO and their derivatives with respect to temperature
in the temperature range of $2 - 300$ K.  The CeMnSbO exhibits a sharp peak at  $T \approx 4.5$ K which, as we demonstrate below,
is due to Ce ordering and Mn internal spin-flop similar to that observed in CeMnAsO\cite{Zhang2015} (spin-flop and spin-reorientation by 90 degrees are used interchangeably throughout). It is interesting to note that  both compounds do not show obvious features in the susceptibilities or their derivatives to indicate a transition from a paramagnetic to a magnetically ordered state.  However, neutron diffraction patterns at low
scattering angles ($2\theta$) show intensity in the forbidden  (100) reflection and significant change in intensities of the (101) and (002) reflections (see Fig.\ \ref{fig:LowTpeaks}) indicating the ordering of Mn sublattice.
\begin{figure} \centering \includegraphics [width = 0.72\linewidth] {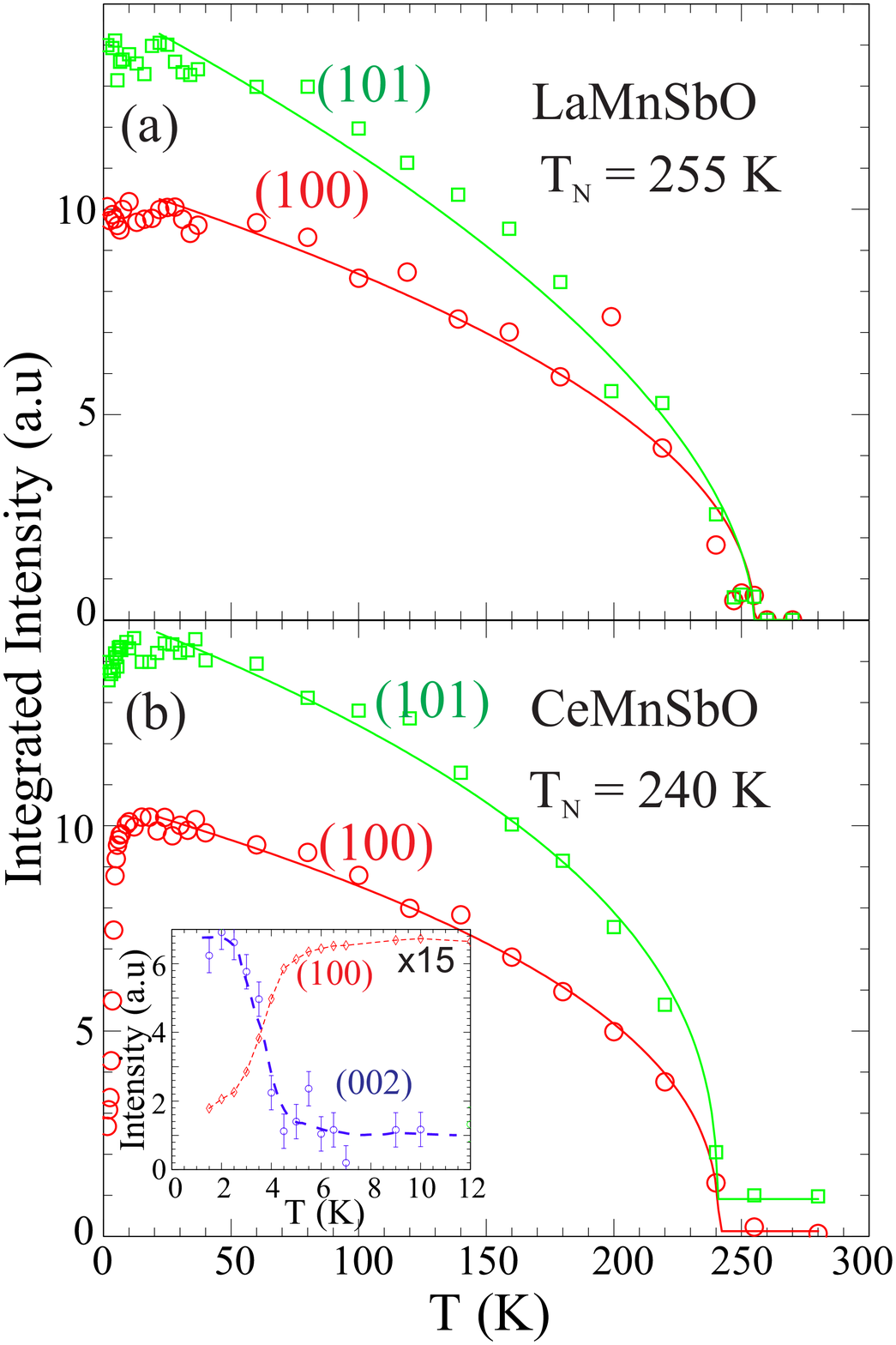}
\caption{(color online) Integrated intensities versus temperature of the (100) and (101) Bragg peaks for (a) CeMnSbO and
(b) LaMnSbO with a fit to a power law (solid lines). At low temperatures the (100) peak of CeMnSbO shows a strong reduction
in intensity around $T_{\rm SR} = 4.5$ K.  }
\label{fig:OrderPara1}
\end{figure}

\begin{figure}\centering\includegraphics [width = 0.88\linewidth]{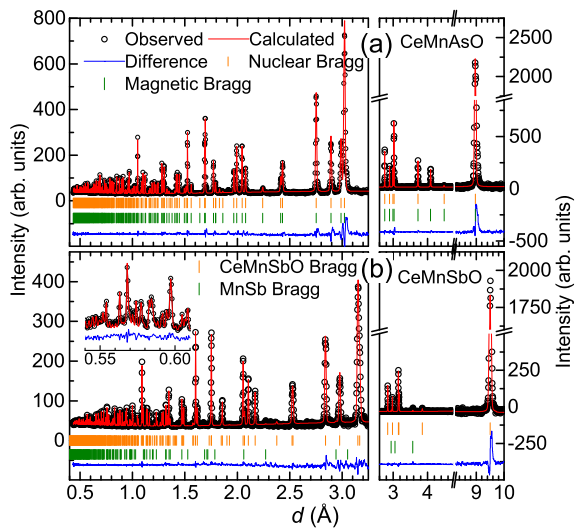}
	\caption{(Colour online)  Observed (measured on POWGEN) and calculated diffraction patterns and their difference for (a) CeMnAsO and (b) CeMnSbO at 300 K. Left and right frame corresponds to the data collected with center wavelengths 1.333~{\AA} and 3.731~{\AA}, respectively. Note that at 300 K CeMnAsO is AFM showing extra reflections that are also accounted for in the calculations. Inset in (b) shows the high quality of the POWGEN data and the calculated fit at the high order reflections region.}
	\label{Refined_Patterns_300K_FullFrame}
\end{figure}

Figure\ \ref{fig:OrderPara1} shows the integrated intensities of the (100) and (101) magnetic Bragg reflections as a function of temperature representing the square of the sublattice magnetic order parameters for both samples.  Fitting the order parameters to a power law, $I(T) = a(1-T/T_{\mathrm{N}})^{2\beta}+b$, yields $T_{\rm N} =255\pm5$ K and  $\beta =0.25(3)$ for LaMnSbO and $T_{\rm N} =240\pm 4$ K and  $\beta =0.24(3)$ in CeMnSbO.  These values are very similar to that obtained for CeMnAsO\cite{Zhang2015,Comment1}.  Such phenomenological power law behavior over the extended range of temperatures with similar exponents has been explained for similar quasi-2D systems\cite{Vaknin1989,Vaknin1990} that possess inplane exchange coupling $J_1$ that  is  significantly  stronger  than the interlayer  one $J_c/J_1 << 1 $\cite{Singh1990}. The quasi-2D behavior extracted from the order parameter is intimately consistent with  the absence of clear AFM signatures in the aforementioned susceptibility.   The absence of clear anomaly at or near  $T_{\rm N}$ in the susceptibility  is  indicative of the two-dimensional nature of the spin system, namely strong  in-plane spin-spin coupling ($J_1$) and very weak inter-planar coupling with likely large but fluctuating spin-correlations above $T_{\rm N}$.  This suggests that the transition temperature $T_{\rm N}$ does not represent the energy scale of nearest-neighbor (NN) coupling $J_1$, i.e., $J_1 >> k_BT_{\mathrm{N}}$.

 As the temperature is lowered below $T_{\mathrm{N}}$ the intensities of the (100) and the (101) for LaMnSbO saturate
 at base temperature (see Fig.\ \ref{fig:LowTpeaks}(a)) indicating that no further magnetic or structural transitions occur below $T_{\mathrm{N}}$.  The Rietveld fit to the NPD pattern at 1.5 K in LaMnSbO and  analysis by SARAh\cite{Wills2000} reveal that the Mn spins form
 a nearest-neighbor antiferromagnetic alignment in the $ab$ plane and ferromagnetic alignment along \textit{c} axis  (C-type AFM order) with moment along the c-axis as shown in
  Fig.\ \ref{fig:LowTpeaks}(c) (left panel). However, as shown in Fig.\ \ref{fig:LowTpeaks} (b) for CeMnSbO below $T \approx 4.5$ K there is a dramatic decrease in
 the intensity of the (100) and a lesser decrease in the (101). In addition, Fig.\ \ref{fig:LowTpeaks} (b) also
 shows a moderate increase in the (002) peak  (more details of its temperature dependence can be found in the inset of Fig.\ \ref{fig:OrderPara1}(b)). Representation analysis by SARAh and detailed Rietveld fit to the whole diffraction patterns at 40 K and 1.5 K indicate that below $\sim 4.5 $ K the Mn spins undergo an internal spin-flop transition in unison from pointing along
 the crystallographic $c$-axis into the plane (main evidence is the behavior of the (100) reflection) while keeping the C-type AFM order.  And simultaneously the AFM ordering of the Ce ions occurs where Ce spins are ferromagnetically aligned in the $ab$ plane and  antiferromagnetically  between planes  (see Fig.\ \ref{fig:LowTpeaks}(c)). The main indicator for the Ce ordering is the abrupt
 increase in the intensity of the (002) reflection (see inset in Fig.\ \ref{fig:OrderPara1}(b) ).
The spin-flop transition is not observed in LaMnSbO involving the nonmagnetic La, but found in CeMnSbO, evidence that the transition is driven by Ce-Mn coupling. The ordering of the Ce
 spins in the {ab} plane indicates a finite single-ion anisotropy for the Ce spin due to its orbital degree of
 freedom ($L =3$) that in this case orients the Ce moment in the basal plane and forces the Mn ions, with the very weak or
 none single-ion-anisotropy ($L=0$) to follow with spin-reorientation to the $ab$-plane via Ce-Mn coupling.

  The average ordered Mn moments of LaMnSbO and CeMnSbO at $\approx 2$ K are found to be 3.45(6) and 3.92(4) $\mu_B$, respectively, both of which are lower than 5
  $\mu_B$ expected for Mn$^{2+}$ in insulators, but large enough to primarily be considered as local-moments.
    The checkerboard-like AFM structure of the C-type order in both compounds below $T_{\rm N}$ suggests that the NN interaction $J_{1}$ is more dominant compared to the NNN  $J_{2}$.
Although their magnetic behaviors  are similar to those in $Ln$MnAsO ($Ln=$ La and Ce),
the Sb-based counterparts differ in three distinct respects: 1). The transition temperatures, $T_{\rm N}= 317$ K\cite{Emery2011}  for LaMnAsO and 347 K for  CeMnAsO \cite {Zhang2015} are  found to be lower by almost 100 K for LaMnSbO and CeMnSbO. 2). The spin-reorientation transition observed at $\approx 35$ K in CeMnAsO is suppressed to $\approx4.5$ K in CeMnSbO. 3). The third magnetic transition at $T^{*}\approx$ 7 K in CeMnAsO\cite{Zhang2015},  possibly related to a collinear-to-noncollinear magnetic structure, is absent inCeMnSbO, implying that the magnetic ground states of CeMnAsO and CeMnSbO are different.
These differences indicate an appreciable effect of the As-by-Sb substitution on the magnetism of $Ln$Mn$Pn$O ($Pn=$ As or Sb). To shed more light on these differences, we have employed POWGEN to systematically study  CeMnAsO and CeMnSbO and compare their structure and magnetism in
detail, taking advantage of the superior instrumental resolution of POWGEN over that of HB1A, and also the wider Q-range (or $d$-spacing range) that it covers. For comparison,
Fig.\ \ref{Refined_Patterns_300K_FullFrame} shows the TOF powder diffraction patterns from  CeMnAsO and CeMnSbO at $T =300$ K.
Both figures also include the best calculated fit to the diffraction patterns that at the large $d$-spacing include magnetic peaks
for CeMnAsO present at 300 K, as this system orders at $T_{\mathrm{N}}= 347$K\cite{Zhang2015}. The high quality of the POWGEN data
and the calculated fit at the high order reflections region are demonstrated in the inset to
Fig.\ \ref{Refined_Patterns_300K_FullFrame}(b).
Table\ \ref{tab:Refined_Parameters} lists the best fit parameters to the diffraction pattern at selected temperatures that represent the various phases of the two systems, in accordance with the magnetic models described above. For both systems we find that the average magnetic moment of Ce is consistently on the order of 1$\mu_B$  whereas the expected value for the free ion is $g(JLS)J = 2.1 \mu_B$ for $4f^1$ configuration of Ce$^{3+}$ ($g(JLS)$ is the Land{\'e} g-factor). This suggests crystal effects play a role in adjusting the actual magnetic moment, an issue that can be further explored by spectroscopic methods.

Also listed in Table\ \ref{tab:Refined_Parameters} are the bond-lengths of As-Mn and Sb-Mn at selected temperatures.
Consistently,  the Sb-Mn bond length is significantly longer than that of the As-Mn bond, which could account for
the significantly lower $T_{\rm N}$ presumably due to a lesser $p-d$ hybridization of $Pn$ and Mn that results in a weaker super-exchange Mn-Mn coupling $J_1$ in SbMn plane.
 The reduced $T_{\rm SR}$ in
CeMnSbO implies a weaker Ce-Ce and Ce-Mn interactions than those in CeMnAsO. Note that in both CeMnAsO and CeMnSbO, each square sheet of O$^{2-}$ ions
is sandwiched between two square sheets of Ce$^{3+}$ ions to form a R-O-R slab and alternates with the As-Mn-As slab along $c$ axis.  Ce$^{3+}$ ions
above and below
the square sheet of the Mn$^{2+}$ form a Mn$_{4}$Ce square pyramid and the superexchange AFM Ce-Ce interaction along $c$-axis is mediated by O atoms.
As shown in the Table\ \ref{tab:Refined_Parameters}, the longer Ce-O bond length in CeMnSbO is probably related
for the weaker AFM super-exchange Ce-Ce interaction in CeMnSbO along $c$-axis.
As for the Ce-Mn interactions, density functional theory (DFT) calculations on CeMnAsO show that they can be complicated due to the presence of multiple interactions such as the Heisenberg, Dzyaloshinskii-Moriya (DM) and biquadratic (BQ) exchange interactions in order of strength.  Thus the  DM (stronger than the BQ) leads to  a noncollinear magnetic structure ground state with the Mn moments orthogonal to those of the Ce moments\cite {Lee2012}. Note that  the NPD is not sufficient in distinguishing the collinear from the noncollinear arrangement between Ce$^{3+}$ and Mn$^{2+}$ moments in CeMnSbO or in CeMnAsO. However  susceptibility measurements and a very weak change in intensity of the (002) magnetic peak at $T^*$ have been observed for CeMnAsO and have been associated with collinear-to-noncllinear transition\cite{Zhang2015}.  Considering the much lower $T_{\rm SR}$ in CeMnSbO, magnetization measurements below $\sim 1$ K and further DFT calculations may shed more light on the issue of the ground state of CeMnSbO. Nevertheless,  the weaker total Ce-Mn interaction in CeMnSbO is most likely related to the larger distance between the two NNs (along $c$-axis) compared to that in the CeMnAsO, as shown in  Table \ \ref{tab:Refined_Parameters}.

\begin{figure}[h]
	\centering
	\includegraphics[scale=0.32]{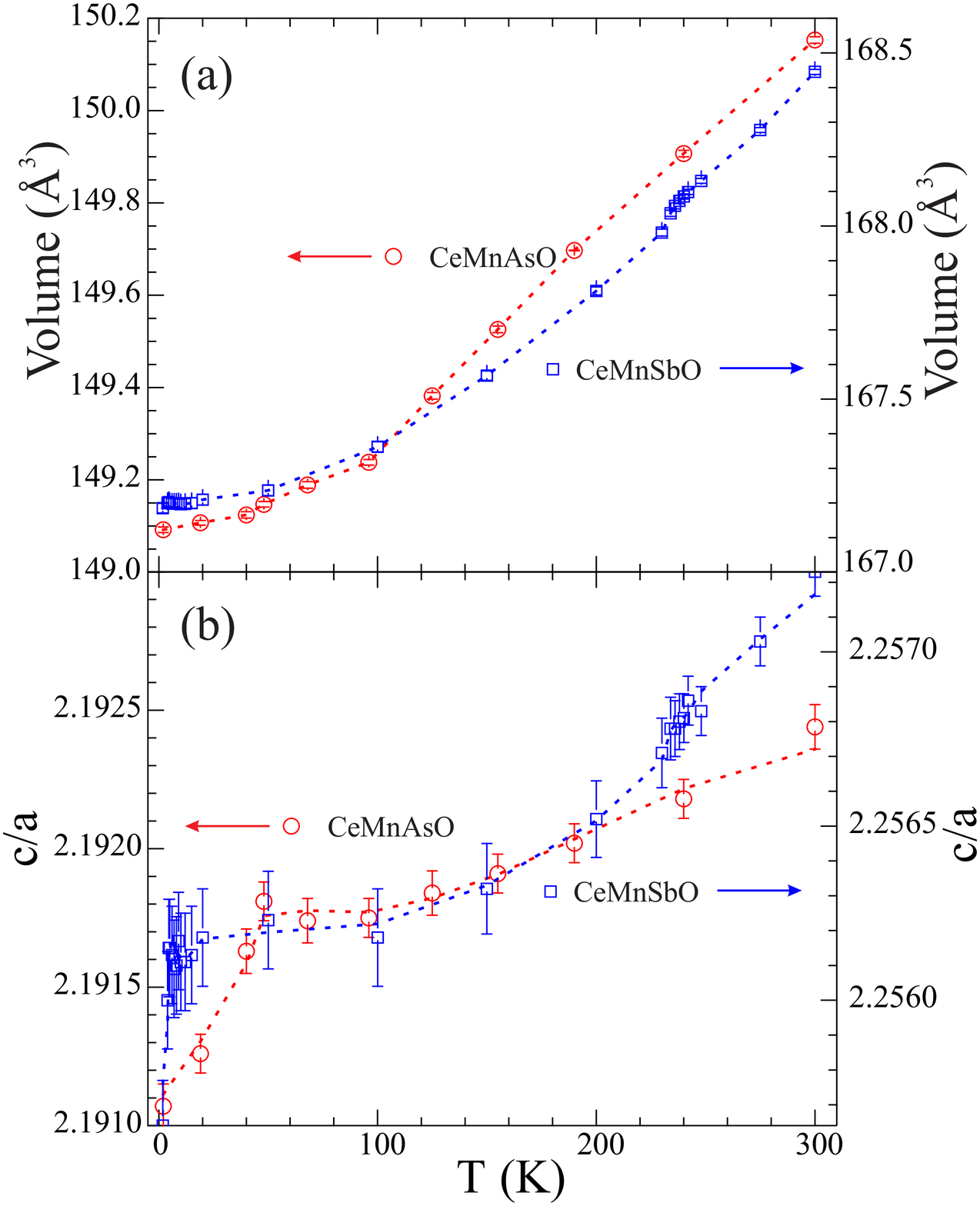}
	\caption{(Colour online) (a) Temperature dependence of cell volumes for  CeMnSbO and CeMnAsO. (b) The $c/a$ ratio of cell parameters
	showing magnetoelastic effect at the Ce ordering, $T_{\rm{Ce}} \approx 4.5$ K for CeMnSbO and $\approx 35$ K for CeMnAsO. Dotted lines are guides to the eye.}
	\label{fig:Cell-a-c}
\end{figure}

The high resolution $d-spacing$ available by the TOF POWGEN
compared to HB1A yields cell parameters with a high relative-accuracy as a function of temperature (i.e., excluding
any systematic error that is temperature independent). As shown in Fig.\ \ref{fig:Cell-a-c}, such high resolution
allows detection of a magnetoelastic effect that manifests itself in an anomaly in the $c/a$ ratio of the cell
parameters as a function of temperature with an onset at $T_{\rm SR}$ of the spin-reorientation transition and Ce-ordering
for CeMnAsO and for CeMnSbO. By contrast the cell volume of both compounds monotonically decreases with decreasing
temperature without anomaly at $T_{\rm{SR}}$.
  \begin{table*}[!]
\centering
\caption{Structural and magnetic data corresponding to simultaneous crystal and magnetic structure refinements of CeMnAsO, CsMnSbO and LaMnSbO at few representative temperatures. All the refinements were carried out with the space group $P4/nmm$ in tetragonal symmetry. Refined results obtained from the measurements at HB1A are indicated by HB1A otherwise the data corresponds to the results obtained from the measurements at POWGEN.}
\label{tab:Refined_Parameters}
\begin{tabular}{C{3cm}c|ccc|ccc|ccc}
 \hline\hline
    &    &\multicolumn{3}{c|}{CeMnSbO}&\multicolumn{3}{c|}{LaMnSbO}& \multicolumn{2}{c}{CeMnAsO}\\ \cline{2-10}  
    &  $T$~(K)      &    300   &    15    &    2     & 272{\tiny (HB1A)} &1.5{\tiny (HB1A)} &   300    &    48    &    2 \\ \hline
    {Unit cell parameters} & &  &&  &&&  & & &\\
  $a~(\AA)$   & & 4.2104(1) &4.2003(1) &4.2003(1)  &4.247(1)  &4.236(1)&4.0914(3) &4.0826(3) &4.0826(1) \\
  $c~(\AA)$   &  &9.5034(2) &9.4774(3) &9.4752(3) &9.572(2) &9.545(2)&8.9701(7) &8.9480(2) &8.9452(3)\\
  $V~(\AA^3)$ & &168.47(1)&167.21(1)&167.17(1)& 172.61(6)&171.25(5)&150.15(1)&149.15(1)&149.09(1) \\ \hline

Atoms (Wyckoff site) &  &   &&& & & & &  &\\
Ce/La~(2c)& $z$                &0.1186(7)&0.1171(6)&0.1195(5)&0.115(2)&0.116(2)&0.1322(6)&0.1322(2)&0.1328(5)\\
     & $B_{iso}~(\AA^{2})$ &0.665(70)&0.308(87)&0.315(84)&0.3(2)	   &0.1(3)&0.17(9)&0.09(3)&0.07(5) \\
     & $m~(\mu_{\rm {B}})$  &     --    &   --    &1.02(4)    &    --    &   --&   --    &   --    &1.08(4) \\  \cline{2-10}

  Mn~(2b) & $B_{iso}~(\AA^{2})$&1.1(2) &0.41(10)&0.322(78)& 2.0(4)    &2.0(5)&0.42(1)&0.195(30)&0.161(61) \\
     &$m~(\mu_{\rm {B}})$    & --      &3.73(4)  &3.92(4)    &   --     &3.45(6) &2.60(5)  &3.41(3)  &4.18(4) \\  \cline{2-10}

As/Sb~(2c)&$z$     			 &0.6827(5)&0.6820(5)&0.6818(4)&0.674(1)&0.675(2)&0.6716(5)&0.6712(2)&0.6704(4)\\
     &$B_{iso}~(\AA^{2})$  &0.823(35)&0.259(66)&0.223(6) & 3.5(4)    &3.1(6)&0.42(8)&0.18(3)&0.17(5)\\  \cline{2-10}

  O~(2a)  &$B_{iso}~(\AA^{2})$ &0.63(4)&0.41(8)&0.37(1)    &0.6(3)    &0.2(4)&0.29(8)&0.20(3)&0.24(5) \\ \hline

  {Bond length}   &  &   \multicolumn{3}{c|}{} & &  &  &  &\\
     & Mn-Sb     &2.729(3)&2.718(3)&2.716(2)&2.700(7)&2.695(8)&2.557(3)&-&- \\
     & Mn-As     &-&-&-&-&-&-&2.5542(9)&2.548(2) \\
     & Ce-O    &2.388(3)&2.386(3)&2.385(2)&-&-&-& 2.359(1)&2.362(2)  \\ \hline\hline
  {Distance of NN ions}   &  &   \multicolumn{3}{c|}{} & &  &  &  &\\

     &  Mn-Mn     &2.9772(1)& 2.9701(1) &2.9701(1)&3.0029(5)&2.9950(3)&-&2.8868(2)&2.8868(1)&\\
     &  Ce-Ce($ab$ plane)    &4.2104(1)&4.2003(1)&4.2003(1)&-&-&-&4.0826(1)&4.0826(1)  \\
     &  Ce-Ce(along $c$)   &3.734(6)& 3.735(5)&3.731(4)&-&-&-&3.7324(2)&3.739(4)  \\
     & Ce-Mn(along $c$)    &4.192(6) & 4.173(5) &4.175(4) &-&-&-& 3.875(4)&3.868(4)  \\\hline\hline

  {Discrepancy factors}  &  &    \multicolumn{3}{c|}{}   &  & & &  &\\
   &$\chi^{2}$      & 1.45 & 1.14 & 1.48 &  2.32 &3.07& 2.68 & 1.88 & 2.19\\
   &$R_p$~(\%)      & 5.04 & 7.00 & 6.2   & 10.9 & 11.9& 5.27 & 5.91 & 6.89\\
   &$R_{wp}~(\%)$   & 7.04 & 10.04& 8.75  & 11.6 & 12.5& 8.48 & 9.46 & 10.8\\
   &$R_{mag}~(\%)$  &  --  & 8.76 & 6.97  &  --  & 13.3& 7.28 & 3.12 & 6.07 \\
  \hline \hline
\end{tabular}
\end{table*}

In addition to inducing spin-flop in $Ln$MnSbO the rare-earth element  has an effect on the lattice. Whereas a T-O transition is observed in PrMnSbO  at 35 K  \cite{Kimber2010}, the tetragonal structure in LaMnSbO and CeMnSbO is
preserved to the lowest temperature ($\sim 2$ K).      Kimber et al. \cite{Kimber2010} have proposed that the T-O structural transition in PrMnSbO is driven
 by the 4f-electron degrees of freedom in connection to multipolar order of 4$f^{2}$ electron of Pr$^{3+}$.  Multipolar ordering of 4$f$ electrons has been discussed in more detail in regard to various transitions in Pr- and Ce-based compounds\cite{Uma1996,Kiss2003,Walker2009,Cameron2015,Matsumura2009}.
 La$^{3+}$ has no 4$f$ electrons and no multipolar ordering.   Ce$^{3+}$ with only one
f-electron (4$f^{1}$) compared to Pr 4$f^{2}$ may exhibit a very weak multipolar ordering that only induces the the $c/a$ anomaly observed at $T_{SR}=$4.5 K as shown in Fig.\ \ref{fig:Cell-a-c}.
It is interesting to point out that a similar scenario as in CeMnSbO has been suggested to the heavy-fermion metal CeB$_{6}$ \cite{Cameron2015}.
 Nuclear magnetic resonance  and resonant x-ray scattering may shed more light  on the different multipolar orders of 4\textit{f} electrons of Ce$^{3+}$ and Pr$^{3+}$ in CeMnSbO and PrMnSbO.

 \section{CONCLUSIONS}
In summary, we report neutron diffraction and magnetization studies on LaMnSbO and CeMnSbO and compare them to other isostructural systems.
The main conclusions from our results are:  1. The Mn$^{2+}$ moments order
in simple AFM checkerboard-like (C-type) structure with the moments along the $c$-axis, with no indication of a transition
in the magnetization as a function of temperature. This and the order-parameter behavior  as a function of temperature,
indicate that the inter-planar coupling is very weak compared to the in-plane coupling, implying a quasi-2D behavior of the magnetic system and that the NN coupling $J_1$ is
likely much larger than $k_BT_{\rm N}$. 2.  The extracted average magnetic moment  at base temperatures ($\sim 4 \mu_B$) is
relatively large and very close to the value expected from a Mn$^{2+}$ ($\sim 5 \mu_B$) with $S = 5/2$.  The small reduction
from the value expected in insulators implies very weak itineracy in the $d$-shell and that to a good approximation in these
systems the Mn spin behaves more like a local-moment.  3. The coupling of Ce-Mn is sufficiently strong to alter the orientation
of the magnetic moments from pointing along the $c$-axis to the $ab$-plane  at $T_{\rm{SR}} = 4.5(5)$ K.  At the same temperature,
the Ce moments undergo AFM ordering at $T_{\rm Ce} = T_{\rm SR}$.  We regard this transition as an internal spin-flop transition
similar to the magnetic field induced spin-flop transition commonly observed in Mn insulators\cite{Toft-Petersen2012}
due to the weak (or none) spin-orbit coupling (angular momentum $L \sim 0$ for Mn$^{2+}$ ion) that tends to result in very
weak single-ion anisotropy such that the spins readily flop by external or internal magnetic fields.
4. The significantly lower Mn ordering temperatures ($T_{\mathrm{N}} \sim 250$ K) of the Sb-  compared to the higher ones for As-based compounds ($T_{\mathrm{N}} \sim 360 K$) implies stronger NN coupling $J_1$ in the latter suggesting a weaker $p-d$ hybridization in Sb-Mn compared to As-Mn.  5. The replacement of As by
Sb  lowers $T_{\rm SR}$ and suppresses the noncolinear transition (seen at  $T^{*}=7$ K in CeMnAsO )   indicating weaker Ce-Ce interactions in in CeMnSbO.  6.   In contrast to PrMnSbO which undergoes a transition to an orthorhombic phase at low temperatures due to Pr$^{3+}$ 4$f^2$ multipolar effects,  the chemical structure of LaMnSbO and CeMnSbO is preserved to the lowest temperature ($\sim 2$ K) suggesting a lesser multipolar effects  in Ce$^{3+}$ 4$f^1$  that only induces a magnetoelastic anomaly observed in the $c/a$ lattice parameters as a function of temperature while as expected, even the magnetoelastic anomaly is absent for the non-magnetic La$^{3+} $(4$f^0$).

\section{Acknowledgments}
Research at Ames Laboratory is supported by the US Department of Energy, Office
of Basic Energy Sciences, Division of Materials Sciences and Engineering under
Contract No. DE-AC02-07CH11358. Use of the High Flux Isotope Reactor and the Spallation Neutron Source at the Oak
Ridge National Laboratory, is supported by the US Department of Energy, Office
of Basic Energy Sciences, Scientific Users Facilities Division.

\end{document}